\documentclass[jkps,preprint,onecolumn,fleqn,showpacs,showkeys]{revtex4}%
\usepackage{graphicx}
\usepackage{amssymb}
\usepackage{amsmath}
\usepackage{bm}
\usepackage{natbib}

\begin{document}

\def\nat{Nature}
\def\prl{Phys. Rev. Lett.}
\def\prb{Phys. Rev. B}
\def\prc{Phys. Rev. C}
\def\prd{Phys. Rev. D}

\def\mnras{Mon. Not. Roy. Astr. Soc.}
\def\apj{Astrophys. J.}
\def\apjl{Astrophys. J. Lett.}
\def\apjs{Astrophys. J. Suppl. Ser.}
\def\aa{Astron. Astrophys.}
\def\aap{Astron. Astrophys.}
\def\actaa{Acta Astronomica}
\def\aapr{Astron. Astrophys. Rev.}
\def\plb{Phys. Lett. B}
\def\pr{Phys. Rev.}
\def\araa{Annual Rev. of Astron. Astrophys.}

\def\pasj{Publications of the Astronomical Society of Japan }
\def\pasp{Publications of the Astronomical Society of the Pacific}
\def\zap{Zeitschrift f{\"u}r Astrophysik}

\def\npa{Nuclear Physics A}
\def\nphysa{Nucl. Phys.}
\def\physrep{Phys. Rep.}

\def\beq{\begin{equation}}
\def\eeq{\end{equation}}
\def\rmd{{\rm d}} 

\def\be{\begin{equation}}
\def\ee{\end{equation}}
\def\bea{\begin{eqnarray}}
\def\eea{\end{eqnarray}}
\setcounter{page}{0}
\title[]{General Relativistic White Dwarfs and Their Astrophysical Implications}

\author{Kuantay \surname{Boshkayev}\footnote{kuantay@icra.it}}
\affiliation{Physical Technical Faculty, Al-Farabi Kazakh National University,\\
Al-Farabi avenue 71, 050038 Almaty, Kazakhstan, \\
ICRANet, Square of Republic 10, I-65122 Pescara, Italy, and}
\author{Jorge A. \surname{Rueda}\footnote{jorge.rueda@icra.it}, Remo \surname{Ruffini}\footnote{ ruffini@icra.it}}
\affiliation{Department of Physics and ICRA, Sapienza University of Rome,\\
Aldo Moro Square 5, I-00185 Rome, Italy\\
ICRANet, Square of Republic 10, I-65122 Pescara, Italy, and}
\author{Ivan \surname{Siutsou}\footnote{siutsou@icranet.org}}
\affiliation{ICRANet, Square of Republic 10, I-65122 Pescara, Italy}

%\footnote{email: kuantay@icra.it, jorge.rueda@icra.it, ruffini@icra.it, siutsou@icranet.org}

\begin{abstract}
We consider applications of general relativistic uniformly-rotating white dwarfs to several astrophysical phenomena related to the spin-up and the spin-down epochs and to delayed  type Ia supernova explosions of super-Chandrasekhar white dwarfs, where we estimate the ``spinning down'' lifetime due to magnetic-dipole braking. In addition, we describe the physical properties of Soft Gamma Repeaters and Anomalous X-Ray Pulsars as massive rapidly-rotating highly-magnetized white dwarfs. Particularly we consider one of the so-called low-magnetic-field magnetars SGR 0418+5729 as a massive rapidly-rotating highly-magnetized white dwarf and give bounds for the mass, radius, moment of inertia, and magnetic field by requiring the general relativistic uniformly-rotating configurations to be stable.
\end{abstract}
\maketitle

\section{Introduction}\label{sec:1}
Rotating white dwarfs (RWDs), depending on their mass, i.e., whether they are sub-Chandrasekhar white dwarfs (WDs) or super-Chandrasekhar WDs, display different behavior. Namely, both uniformly and differentially rotating super-Chandrasekhar white dwarfs (SCWDs) spin-up by angular momentum loss whereas sub-Chandrasekhar WDs only spin-down by angular momentum loss. We should mention that the spin-up of rapidly-rotating stars was first described by Shapiro et al. \cite{shapiro90} and later by Geroyannis and Papasotiriou \cite{2000ApJ...534..359G}. In both Refs. \cite{shapiro90} and \cite{2000ApJ...534..359G} the authors performed computations in classical physics without taking into account the effects of general relativity (GR) although GR is very crucial in investigating the stability of RWDs \cite{shapirobook}.

Several scenarios for obtaining SCWDs exist: a single degenerate scenario \cite{1973ApJ...186.1007W,1982ApJ...253..798N,2004MNRAS.350.1301H}, where a WD grows in mass through accretion from a non-degenerate stellar companion; a double degenerate scenario \cite{1984ApJS...54..335I,1984ApJ...277..355W,2010ApJ...722L.157V}, where two WDs merge after losing energy and angular momentum through the radiation of gravitational waves;  and a core degenerate scenario \cite{2011MNRAS.417.1466K}, where the merger occurs in a common envelop with a massive asymptotic branch star. According to Ilkov and Soker \cite{ilkov2012}, SCWDs explode as type Ia supernovae on a spin-down time scale $\tau_B$ due to magnetic-dipole braking. The characteristic time scale of a SCWD before supernova explosion has been estimated by Ilkov and Soker \cite{ilkov2012} to be $10^7 \lesssim \tau_B \lesssim 10^{10}$ yr for magnetic fields in the range $10^6 \lesssim B \lesssim 10^{8}$ G.

Moreover, following and extending the ideas of Morini et al. \cite{1988ApJ...333..777M} and Paczynski \cite{paczynski90}, Malheiro et al. \cite{M2012} proposed the model of massive highly-magnetized fast RWDs. According to Malheiro et al., unlike the widely accepted the magnetar model \cite{mereghetti08, rea10}, the basic physical properties of Soft Gamma Repeaters and Anomalous X-Ray Pulsars (SGRs and AXPs) can be well explained within the WD model. The advantages and the drawbacks of both models are discussed in Ref. \cite{M2012}.

In our recent work \cite{2013ApJ...762..117B}, we computed general relativistic configurations of uniformly RWDs within Hartle's formalism \cite{1967ApJ...150.1005H}. We used the relativistic Feynman-Metropolis-Teller equation of state \cite{2011PhRvC..83d5805R} for WD matter, which generalizes the traditionally-used equation of state of Salpeter \cite{1961ApJ...134..669S}. The stability of rotating WDs was analyzed taking into account the mass-shedding limit, inverse $\beta$-decay instability, and secular axisymmetric instability, with the last being determined by using the turning point method of Friedman et al. \cite{1988ApJ...325..722F}. It has been  shown there that RWDs can be stable up to rotation periods of $\sim 0.28$ s (see Ref. \cite{2013ApJ...762..117B} and Sec.~\ref{sec:3} for details). This range of stable rotation periods for WDs amply covers the observed rotation rates of SGRs and AXPs, $P\sim (2$--$12)$ s.
The minimum rotation period $P_{min}$ of WDs is obtained for a configuration rotating along the Keplerian sequence at the critical inverse $\beta$-decay density; namely, this is the configuration lying at the crossing point between the mass-shedding and the inverse $\beta$-decay boundaries. The numerical values of the minimum rotation period $P_{min}\approx 0.3,0.5,0.7$ and $2.2$ s and the maximum masses $M_{max}^{J\neq0}\approx 1.500, 1.474, 1.467, $ and $1.202$ solar mass were found for helium (He), carbon (C), oxygen (O), and iron (Fe) WDs, respectively \cite{2013ApJ...762..117B}.

Recently, Rueda et al. \cite{2013ApJ...772L..24R} studied the possibility that the peculiar AXP 4U 0141+61 is a massive, rapidly-rotating, highly-magnetized white dwarf and explored the viability of this object being the result of the coalescence of a binary white dwarf. Specifically, from its observed rotational velocity, the bounds for the mass, radius, moment of inertia, etc. were derived (see Ref. \cite{2013ApJ...772L..24R} for details).

In this work, we consider the astrophysical implications of rotating WDs based on the results of Boshkayev et al. \cite{2013ApJ...762..117B}. Particularly, in order to investigate the spin-up and the spin-down of both sub- and super-Chandrasekhar WDs, unlike Shapiro et al. \cite{shapiro90} and Geroyannis and Papasotiriou \cite{2000ApJ...534..359G}, we perform our computations in GR. We also perform computations analogous to those of Ilkov and Socker \cite{ilkov2012} to estimate the spin-down time scale $\tau_B$ due to magnetic dipole braking in GR by relaxing the constancy of the radius and the moment of inertia of a RWD. Eventually, fulfilling all the stability criteria for RWDs in GR on the basis of the WD model of Malheiro et al. \cite{M2012}, we consider one of the so-called low-magnetic-field magnetars SGR 0418+5729 and give bounds for the mass, radius, moment of inertia, and magnetic field.
Our paper is organized as follows: In Section \ref{sec:2}, we discuss spin-up and spin-down of rotating SCWDs, in Section \ref{sec:3}, we consider delayed supernova explosions of SCWDs, and in Section \ref{sec:4}, we calculate the main physical parameters of SGR 0418+5729 within the massive rapidly-rotating highly-magnetized white dwarf model. Finally, in Section \ref{sec:5}, we summarize our main results, discuss their significance, and draw our conclusions.
%
%%%%%%%%%%%%%%%%%%%%%%%%%%%%%%%%%%%%%%%%%%%%%%%%%%%%%%%%%%%%%%%%%%%%%%%%%
%%%%%%%%%%%%%%%%%%%%%%%%%%%%%%%%%%%%%%%%%%%%%%%%%%%%%%%%%%%%%%%%%%%%%%%%%
\section{Spin-up and spin-down epochs of rotating Super-Chandrasekhar White Dwarfs}\label{sec:2}
\begin{figure}
\centering
\includegraphics[width=1\columnwidth,clip]{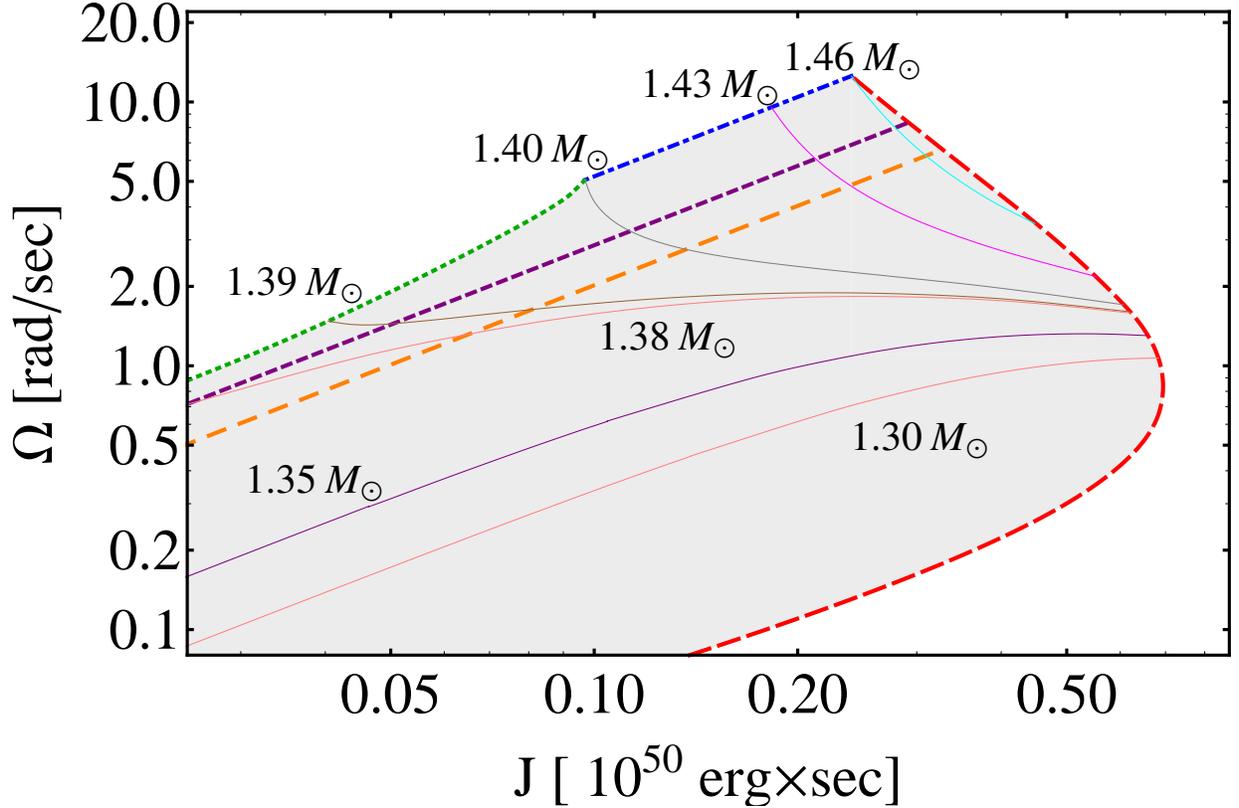} 
\caption{Contours of constant rest-mass in the $\Omega-J$ plane; RWDs that evolve along a track with $\partial \Omega/\partial J>0$ spin-down by loosing angular momentum while the ones with $\partial \Omega/\partial J<0$  spin-up.}\label{fig:consOmJ2}
\end{figure}

%%%%%%%%%%%%%%%%%%%%%%%%%%%%%%%%%%%%%%%%%%%%%%%%%%%%%%%%%%%%%%%%%%%%%%%%%
%%%%%%%%%%%%%%%%%%%%%%%%%%%%%%%%%%%%%%%%%%%%%%%%%%%%%%%%%%%%%%%%%%%%%%%%%
At constant rest-mass $M_0$, entropy $S$ and chemical composition $(Z,A)$, the spin evolution of a RWD is given by (see Ref. \cite{shapiro90} for details)
\begin{equation}
\dot{\Omega} = \frac{\dot{E}}{\Omega} \left( \frac{\partial \Omega}{\partial J} \right)_{M_0,S,Z,A}\, ,
\end{equation}
where $\dot{\Omega} \equiv d\Omega/dt$, $\dot{E} \equiv dE/dt$, $\Omega$ is the angular velocity, $E$ is the energy, and $J$ is the angular momentum of the star.
\begin{figure}
\centering
\includegraphics[width=1\columnwidth,clip]{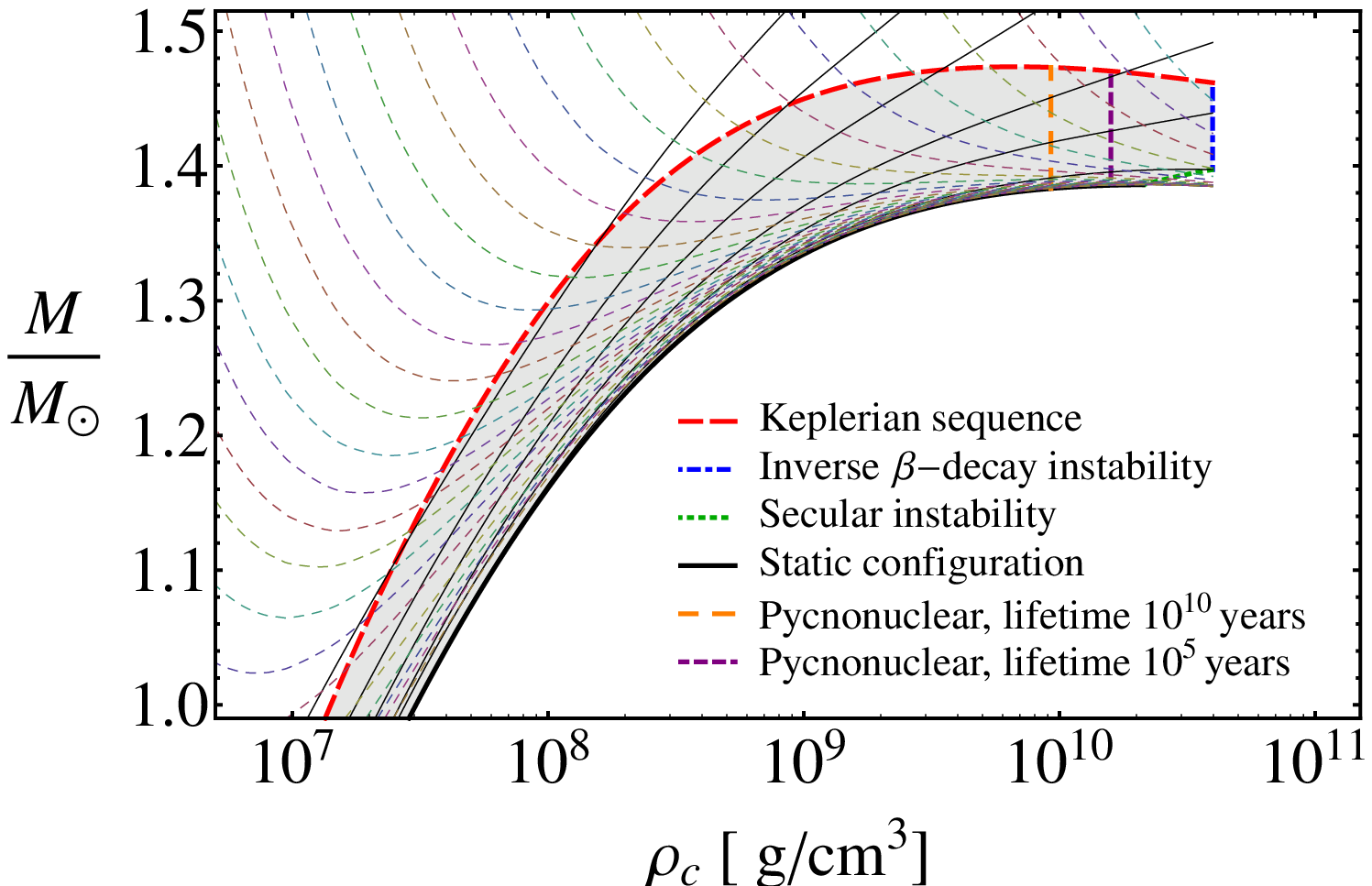} 
\caption{Mass versus the central density for $^{12}$C RWDs. The solid black curves correspond to $J$=constant sequences, where the static case $J=0$ is the thickest one. The colored thin-dashed curves correspond to $\Omega$=constant sequences. The Keplerian sequence is the red thick-dashed curve, the blue thick-dotted-dashed curve is the inverse $\beta$-decay instability boundary, and the green-thick dotted curve is the axisymmetric secular instability boundary.}\label{fig:consOmJ}
\end{figure}
Thus, if a RWD is losing energy by some mechanism during its evolution, that is, $\dot{E}<0$, the change in the angular velocity $\Omega$ with time depends on the sign of $\partial \Omega/\partial J$; RWDs that evolve along a track with $\partial \Omega/\partial J>0$ will spin-down ($\dot{\Omega}<0$), and the ones following tracks with $\partial \Omega/\partial J<0$ will spin-up ($\dot{\Omega}>0$).
In Fig.~\ref{fig:consOmJ}, we show $\Omega=$constant and $J=$constant sequences in the mass-central density diagram, and in Fig.~\ref{fig:consOmJ2}, we show contours of constant rest-mass in the $\Omega-J$ plane.
The sign of $\partial\Omega/\partial J$ can be analyzed from Fig.~\ref{fig:consOmJ} by joining two consecutive $J=$ constant sequences with an horizontal line and taking into account that $J$ decreases from left to right and from top to bottom. The angular velocity $\Omega$, instead, decreases from right to left and from top to bottom for SCWDs, and for sub-Chandrasekhar WDs, it decreases from left to right and from top to bottom. We note that, in the SCWD region, $\Omega=$ constant sequences satisfy $\partial \Omega/\partial \rho_c<0$ while in the sub-Chandrasekhar region, both $\partial \Omega/\partial \rho_c<0$ and $\partial \Omega/\partial \rho_c>0$ appear (see minima). SCWDs can only either spin-up by angular momentum loss or spin-down by gaining angular momentum. In the latter case, the RWD becomes decompressed with time, increasing its radius and moment of inertia: thus, SCWDs following this evolution track will end at the mass-shedding limit (see Fig.~\ref{fig:consOmJ}).

Some evolutionary tracks of sub-Chandrasekhar WDs and SCWDs are shown in Fig.~\ref{fig:consOmJ2}. It is appropriate to recall here that Shapiro et al. \cite{shapiro90} showed that spin-up behavior by angular momentum loss occurs for rapidly-rotating Newtonian polytropes if the polytropic index is very close to $n=3$, namely, for an adiabatic index $\Gamma \approx 4/3$. Geroyannis and Papasotiriou \cite{2000ApJ...534..359G} explicitly showed that those conditions were achieved only by super-Chandrasekhar polytropes.
Besides the corroboration of the above known result for SCWDs in the general relativistic case, we report here the presence of minima $\partial \Omega/\partial \rho_c=0$ for some sub-Chandrasekhar masses (see, e.g., the evolution track of the RWD with $M=1.38 M_\odot$ in Fig.~\ref{fig:consOmJ2}), which raises the possibility that sub-Chandrasekhar WDs can experience, by angular momentum loss, not only the intuitive spin-down evolution but also spin-up epochs.
%%%%%%%%%%%%%%%%%%%%%%%%%%%%%%%%%%%%%%%%%%%%%%%%%%%%%%%%%%%%%%%%%%%%%%%%%%%%%%%%%%%%%%%%%%%%%%%%%%%%%%%%%%%%%%%%%%%%%%%%%%%%%%%%%%%%%%%%%%%%%%%%%%%%%%%%%%%%%%%%%%%%%%%%%%%%%%%%%%%%%%%%%%%%%%%%%%%%%%%%%%%%%%%%%%%%%%%%%%%%%%%%%%%%%%%%%
\section{Delayed Supernova Explosion of Super Chandrasekhar White Dwarfs}\label{sec:3}
%%%%%%%%%%%%%%%%%%%%%%%%%%%%%%%%%%%%%%%%%%%%%%%%%%%%%%%%%%%%%%%%%%%%%%%%%%%%%%%%%%%%%%%%%%%%%%%%%%%%%%%%%%%%%%%%%%%%%%%%%%%%%%%%%%%%%%%%%%%%%%%%%%%%%%%%%%%%%%%%%%%%%%%%%%%%%%%%%%%%%%%%%%%%%%%%%%%%%%%%%%%%%%%%%%%%%%%%%%%%%%%%%%%%%%%%
The majority of  observed magnetic WDs are massive, for instance, REJ 0317-853 with $M \sim 1.35 M_\odot$ and $B\sim (1.7$--$6.6)\times 10^8$ G (see, e.g., Refs. \cite{1995MNRAS.277..971B} and \cite{2010A&A...524A..36K}), PG 1658+441 with $M \sim 1.31 M_\odot$ and $B\sim 2.3\times 10^6$ G (see, e.g., Refs. \cite{1983ApJ...264..262L} and \cite{1992ApJ...394..603S}), and PG 1031+234 with the highest magnetic field $\sim 10^9$ G (see, e.g., Refs. \cite{1986ApJ...309..218S} and \cite{2009A&A...506.1341K}). However, they are generally found to be slow-rotators (see, e.g., Ref. \cite{2000PASP..112..873W}). Recently Garcia-Berro et al. \cite{enrique2012} have shown that such magnetic WDs can be, indeed, the result of the merger of double degenerate binaries; the misalignment of the final magnetic dipole moment of the newly-born RWD with the rotation axis of the star depends on the difference between the masses of the WD components of the binary.
The magnetic braking of SCWDs has been recently invoked as a possible mechanism to explain the delayed time distribution of type Ia supernovae (SNe) (see Ref. \cite{ilkov2012} for details): a type Ia SN explosion is delayed for a time typical of the spin-down time scale $\tau_B$ due to magnetic braking, providing the result of the merging process of a WD binary system is a magnetic SCWD rather than a sub-Chandrasekhar one. The characteristic time scale $\tau_B$ of the SCWD has been estimated to be $10^7 \lesssim \tau_B \lesssim 10^{10}$ yr for magnetic fields in the range $10^6 \lesssim B \lesssim 10^{8}$ G.

A constant moment of inertia $\sim 10^{49}$ g cm$^2$ and a fixed critical (maximum) rotation angular velocity,
\begin{equation}
\Omega_{\rm crit}\sim 0.7 \Omega^{J=0}_{\rm K}=0.7 \sqrt{\frac{G M^{J=0}}{R^3_{M^{J=0}}}}\, ,
\end{equation}
have been adopted in Ref. \cite{ilkov2012}.
It is important to recall here that SCWDs spin-up by angular momentum loss; therefore, the reference to a ``spin-down'' time scale for them is just historical. SCWDs then evolve toward the mass-shedding limit, which determines, in this case, the critical angular velocity for rotational instability (see Fig.~\ref{fig:consOmJ2}).
If we express $\Omega_{K}^{J\neq0}$ in terms of $\Omega_{K}^{J=0}$, taking into account the values of the dimensionless angular momentum $j$ and the dimensionless quadrupole moment $q$ from the numerical integration, we find for RWDs, that $\Omega_{K}^{J\neq0}=\sigma\Omega_{K}^{J=0}$, where the coefficient $\sigma$ varies in the interval [0.78, 0.75] in the range of central densities $[10^5, 10^{11}]$ g cm$^{-3}$ (see Boshkayev et al.  \cite{2013ApJ...762..117B} for details). It is important to mention that the above range of $\sigma$ remains approximately the same independently of the chemical composition of the WD.
\begin{figure}
\centering
\includegraphics[width=1\columnwidth,clip]{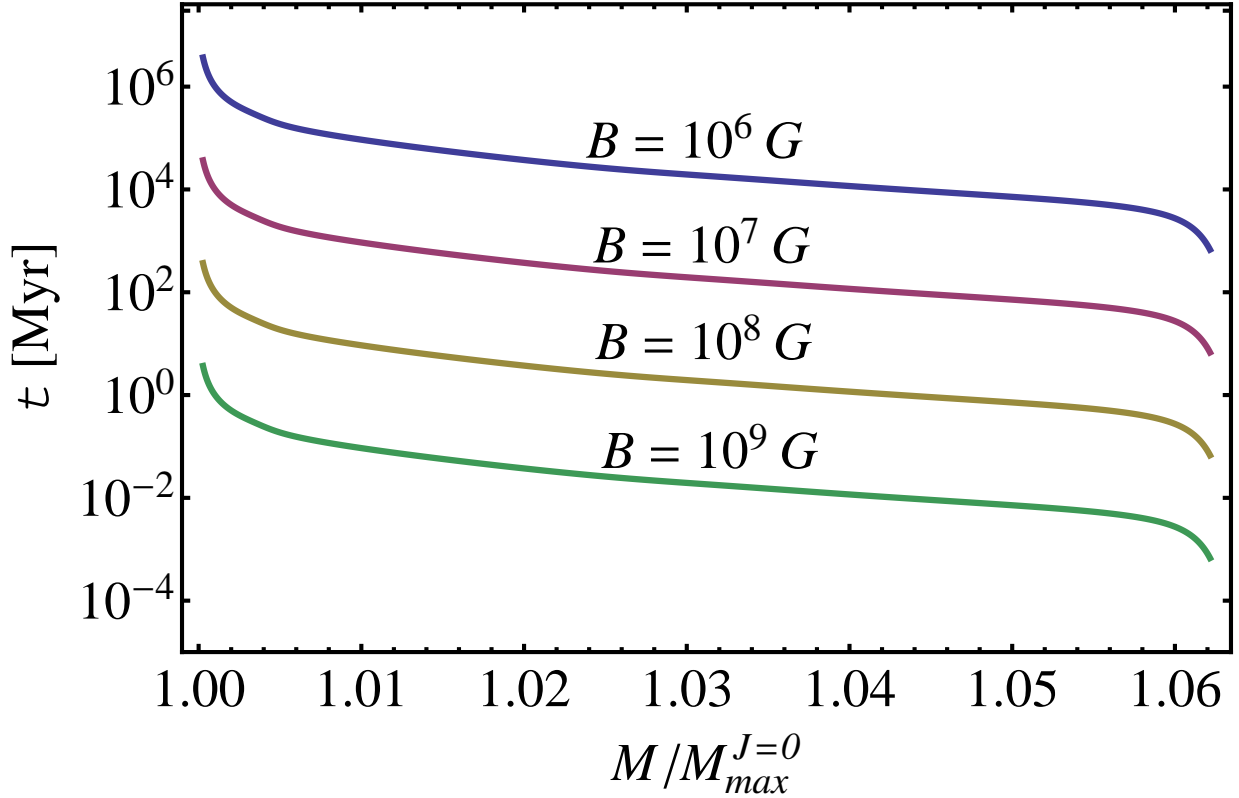} 
\caption{Characteristic life time $t$ in Myr versus WD mass in units of $M^{J=0}_{max}$ for $^{12}$C RWDs. The magnetic field $B$ is in Gauss.}\label{fig:lifetime}
\end{figure}
Furthermore, as we have shown, the evolution track followed by a SCWD depends strongly on the initial conditions of mass and angular momentum, as well as on the chemical composition and the evolution of the moment of inertia. Clearly, the assumption of fixed moment of inertia, $I\sim 10^{49}$ g cm$^2$, leads to a spin-down time scale that depends only on the magnetic field's strength. A detailed computation will lead to a strong dependence on the mass of the SCWD, resulting in a two-parameter family of delayed times $\tau_B(M,B)$.

To the above, we have performed similar analyses in order to estimate the characteristic time scale for the realistic equation of state of WDs presented in Ref. \cite{RotD2011} by relaxing the constancy of the moment of inertia, radius and other parameters of RWDs. Indeed we have shown here that all parameters are functions of the central density and the angular velocity (rotation period): 
\begin{equation}\label{eq:lifetime}
\begin{split}
\tau_B(M,B)=t=-\frac{3c^3}{2B^2}\int_{J_{max}}^{J_{min}}{\frac{1}{R^6}\frac{dJ}{\Omega^3}},\quad
\\ \quad R=R(\rho,J), \quad \Omega=\Omega(\rho,J),\quad
\end{split}
\end{equation}
where $J$, $\Omega$ and $R$ are calculated along the specific constant-rest-mass sequences shown in  Fig.~\ref{fig:consOmJ2}. Hence, we have performed more refined analyses by taking into consideration all the stability criteria, except for the pycnonuclear instabilities, for the sake of generality.

The characteristic time $t$ or $\tau$ versus WD mass in units of $M^{J=0}_{max}$ for $^{12}$C RWDs is shown in Fig.~\ref{fig:lifetime}, where we can see that the higher the magnetic field, the shorter the lifetime of the rotating super-Chandrasekhar WD. Correspondingly, a more massive WD will have a shorter life span and vice versa. Interestingly, the time scales shown in Fig.~\ref{fig:lifetime} are consistent with those of Ilkov and Socker \cite{ilkov2012} and K{\"u}lebi et al. \cite{2013MNRAS.431.2778K}.
%%%%%%%%%%%%%%%%%%%%%%%%%%%%%%%%%%%%%%%%%%%%%%%%%%%%%%%%%%%%%%%%%%%%%%%%%%%%%%%%%%%%%%%%%%%%%%%%%%%%%%%%%%%%%%%%%%%%%%%%%%%%%%%%%%%%%%%%%%%%%%%%%%%%%%%%%%%%%%%%%%%%%%%%%%%%%%%%%%%%%%%%%%%%%%%%%%%%%%%%%%%%%%%%%%%%%%%%%%%%%%%%%%%%%%%%
\section{Soft Gamma-Ray-Repeaters and Anomalous X-ray Pulsars as Massive Rapidly-Rotating Highly-Magnetized White Dwarfs: the case of SGR 0418+5729}\label{sec:4}

SGRs and AXPs are a class of compact objects that show interesting observational properties not observed in ordinary pulsars (see, e.g., Ref. \cite{mereghetti08}): rotational periods in the range $P\sim 2$--$12$ s, a narrow range with respect to the wide range of ordinary pulsars $P\sim 0.001$--$10$ s; spin-down rates $\dot{P} \sim 10^{-13}$--$10^{-10}$, larger than ordinary those of pulsars $\dot{P} \sim 10^{-15}$; strong outbursts of energies, $\sim 10^{41}$--$10^{43}$ erg; and for the case of SGRs, giant flares of even larger energies of $\sim 10^{44}$--$10^{47}$ erg.
Here, we describe one of the so-called {\it low-magnetic-field magnetars}, SGR 0418+5729, as a massive rapidly-rotating highly-magnetized WD and infer from theory basic physical parameter of that source. In doing so, we extend the work of Malheiro et al. \cite{M2012} by using the precise WD parameters recently obtained by Boshkayev et al. \cite{2013ApJ...762..117B} for general relativistic uniformly-rotating WDs.

The loss of rotational energy associated with the spin-down of the WD is given by
\begin{equation}\label{eq:Edot}
\begin{split}
\dot{E}_{\rm rot} = -4 \pi^2 I \frac{\dot{P}}{P^3}\qquad\qquad\qquad
\\= -3.95\times 10^{50} I_{49} \frac{\dot{P}}{P^3}\quad {\rm erg s}^{-1}\, ,
\end{split}
\end{equation}
where $I_{49}$ is the moment of inertia of the WD in units of $10^{49}$ g cm$^2$. This rotational energy loss amply justifies the steady X-ray emission of all SGRs and AXPs (see Ref. \cite{M2012} for details).
\begin{figure}
\centering
\includegraphics[width=1\columnwidth,clip]{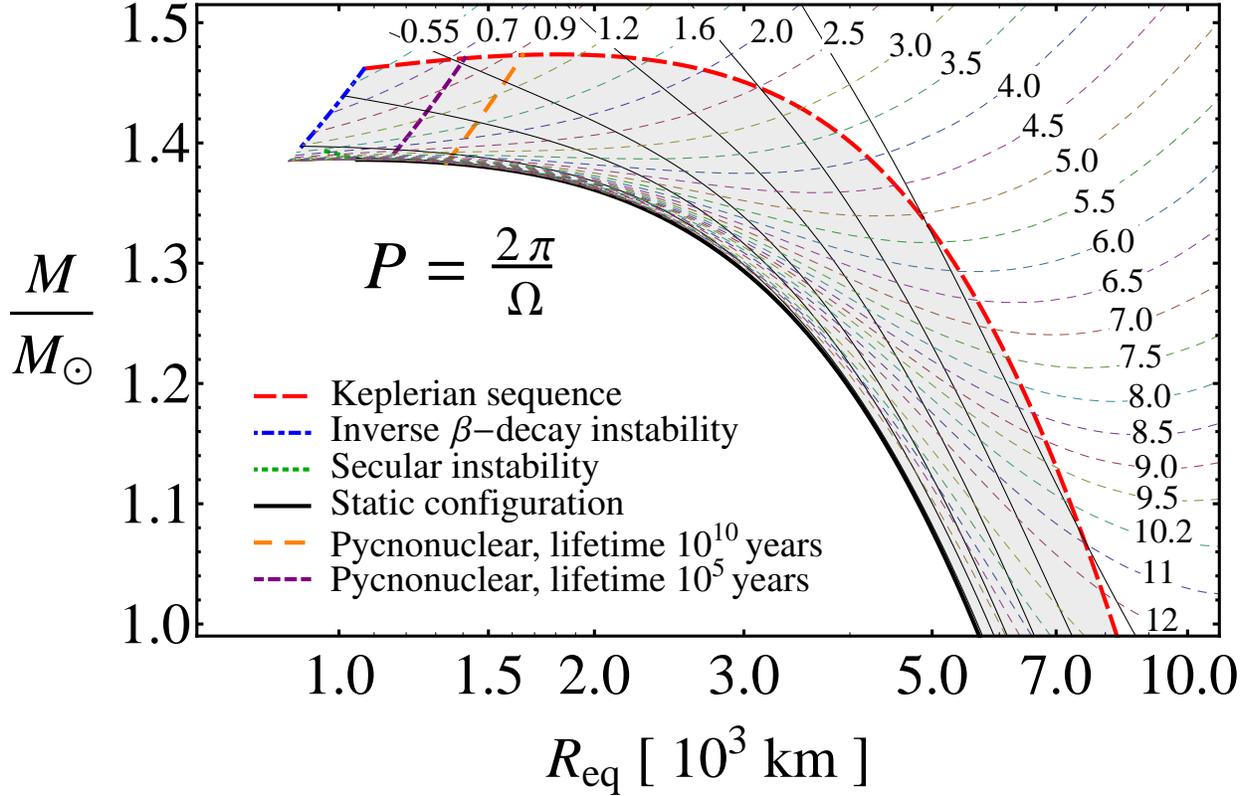}
\caption{Mass versus equatorial radius of rotating carbon WDs. The solid black curves correspond to $J$=constant sequences, where the static case $J=0$ is the thickest one. The colored thin-dashed curves correspond to $\Omega$=constant sequences. The Keplerian sequence is the red thick-dashed curve, the blue thick-dotted-dashed curve is the inverse $\beta$ instability boundary, and the green thick-dotted curve is the axisymmetric instability line. The gray-shaded region is the stability region of rotating white dwarfs \cite{2013ApJ...762..117B}. The numbers show the period $P$ in seconds.}\label{fig:consOmJ3}
\end{figure}
The upper limit on the magnetic field (see, e.g., Ref. \cite{ferrari69}) obtained by requiring that the rotational energy loss due to the dipole field be smaller than the electromagnetic emission of the magnetic dipole is given by
\begin{equation}\label{eq:Bmax}
\begin{split}
B=\left(\frac{3 c^3}{8 \pi^2} \frac{I}{\bar{R}^6} P \dot{P} \right)^{1/2}\qquad\qquad
\\=3.2\times 10^{15} \left(\frac{I_{49}}{\bar{R}^6_8}P \dot{P} \right)^{1/2} {\rm G}\, ,
\end{split}
\end{equation}
where $\bar{R}_8$ is the mean radius of the WD in units of $10^8$ cm. The mean radius is given by $\bar{R}=(2 R_{eq}+R_p)/3$ (see, e.g., Ref. \cite{1968ApJ...153..807H}) with $R_{eq}$ and $R_p$ being the equatorial and the polar radius of the star, respectively.
Clearly, the specific values of the rotational energy loss and the magnetic field depend on the observed parameters, such as $P$ and $\dot{P}$, and on the model parameters, such as the mass, moment of inertia, and mean radius of the rotating WD.

From Fig.~\ref{fig:consOmJ3}, every $\Omega$=constant or equivalently $P$=constant sequence can be seen to intersect the stability region of general relativistic uniformly-RWDs ($M$-$R_{eq}$ curves inside the shaded region) at two points. These two points determine the minimum (maximum) and maximum (minimum) $M_{min,max}$($R^{max,min}_{eq}$), respectively, for the stability of a WD with the given rotation angular velocity $\Omega=2 \pi/P$. Associated with the boundary values $M_{min,max}$ and $R^{max,min}_{eq}$, we can obtain the corresponding bounds for the moment of inertia of the WD, $I_{max,min}$, respectively (see Ref.  \cite{2013A&A...555A.151B} for details). 
\begin{table*}
\centering
\caption{Bounds for the mass $M$ (in units of $M_\odot$), equatorial $R_{eq}$ and mean $\bar{R}$ radius (in units of $10^8$ cm),  moment of inertia $I$, and surface magnetic field $B$ of SGR 0418+5729. $I_{48}$ and $I_{50}$ are the moments of inertia in units of $10^{48}$ and $10^{50}$ g cm$^2$, respectively.}
{\scriptsize
\begin{tabular}{c c c c c c c c c c c}
Comp. & $M_{min}$ & $M_{max}$ & $R^{min}_{eq}$ & $R^{max}_{eq}$ & $\bar{R}_{min}$ & $\bar{R}_{max}$ &
$I^{min}_{48}$ & $I^{max}_{50}$ & $B_{min} (10^7 {\rm G})$ & $B_{max} (10^8 {\rm G})$\\
\hline He & 1.18 & 1.41 & 1.16 & 6.88 & 1.15 & 6.24 & 3.59 & 1.48 & 1.18 & 2.90\\
C & 1.15 & 1.39 & 1.05 & 6.82 & 1.05 & 6.18 & 2.86 & 1.42 & 1.19 & 3.49\\
O & 1.14 & 1.38 & 1.08 & 6.80 & 1.08 & 6.15 & 3.05 & 1.96 & 1.42 & 3.30\\
Fe & 0.92 & 1.11 & 2.21 & 6.36 & 2.21 & 5.75 & 12.9 & 1.01 & 1.25 & 0.80\\
\hline
\end{tabular}
}
\label{tab:SGR0418}
\end{table*}
%%%%%%%%%%%%%%%%%%%%%%%%%%%%%%%%%%%%%%%%%%%%%%%%%%%%%%%%%%%%%%%%%%
%%%%%%%%%%%%%%%%%%%%%%%%%%%%%%%%%%%%%%%%%%%%%%%%%%%%%%%%%%%%%%%%%%
Thus, to calculate all the bounds, we need to know only the rotational period $P$ and the spin-down rate $\dot{P}$ from observations of the source considered. 

SGR 0418+5729 has a rotational period of $P=9.08$ s, and the upper limit of the spin-down rate $\dot{P} < 6.0 \times 10^{-15}$ was obtained by Rea et al. \cite{rea10}. The corresponding rotation angular velocity of the source is $\Omega=2\pi/P=0.69$ rad s$^{-1}$. We show in Table \ref{tab:SGR0418} bounds for the mass, equatorial radius, mean radius, and moment of inertia of SGR 0418+5729 (see Ref. \cite{2013A&A...555A.151B} for details) obtained by the requiring rotational stability of the rotating WD for selected chemical compositions. Although we derived the bounds of parameters for the specific source SGR 0418+5729, the results of Table \ref{tab:SGR0418} are consistent with the observed values of the mass and the magnetic field for other massive magnetic WDs \cite{1995MNRAS.277..971B,2010A&A...524A..36K,1983ApJ...264..262L,1992ApJ...394..603S,1986ApJ...309..218S,2009A&A...506.1341K}.
%%%%%%%%%%%%%%%%%%%%%%%%%%%%%%%%%%%%%%%%%%%%%%%%%%%%%%%%%%%%%%%%%%%
%%%%%%%%%%%%%%%%%%%%%%%%%%%%%%%%%%%%%%%%%%%%%%%%%%%%%%%%%%%%%%%%%%
\section{Concluding Remarks}
\label{sec:5}
%%%%%%%%%%%%%%%%%%%%%%%%%%%%%%%%%%%%%%%%%%%%%%%%%%%%%%%%%%%%%%%%%%
In this work, we showed that all stable uniformly-RWDs can exist only in the stability region. Hence, we assume that we have various WDs at different times in their evolution without involving the details and the routes of their entire evolution.

We also showed that, by losing angular momentum, sub-Chandrasekhar RWDs can experience both spin-up and spin-down epochs while SCWDs can only experience spin-up epochs. These results are particularly important for the evolution of WDs whose masses approach, either from above or from below, the maximum non-rotating mass. 

Knowing the actual values of the masses, radii, and moments of inertia of massive RWDs, we have computed the delay times in models of type Ia SN explosions.

The recent observations of SGR 0418+5729 (see Ref. \cite{rea10}), $P=9.08$ and $\dot{P} < 6.0 \times 10^{-15}$,
challenge the description of this source as a ultramagnetized NS of the magnetar model. Based on the recent work of Malheiro et al. \cite{M2012}, we have shown here that, instead, SGR 0418+5729 is in full agreement with a description based on a massive rapidly-rotating highly-magnetic WD. 

From the analysis of the rotational stability of the WD by using the results of Boshkayev et al. \cite{2013ApJ...762..117B}, we have predicted the WD parameters: in particular, bounds for the mass, radius, moment of inertia, and magnetic field of SGR 0418+5729.

Throughout the paper, we performed computations in GR by using the Hartle formalism for uniformly-rotating configurations. We considered mainly WDs consisting of carbon, although the typical white dwarfs are known not to  consist of a pure chemical element, but a mixture of elements such as carbon, oxygen, neon, magnesium, etc. It would be interesting to consider the chemical profiles of Renedo et al. \cite{2010ApJ...717..183R} in our future investigations.
\begin{acknowledgments}
The authors wish to thank the organizers and the speakers of the 13th Italian-Korean Symposium for the excellent conference and acknowledge the hospitality at Asia Pacific Center for Theoretical Physics (APCTP) where part of this work was done. K.B. acknowledges ICRANet for its hospitality and support. The authors thank the anonymous referee for useful recommendations, comments and suggestions which helped to improve the paper.  
\end{acknowledgments}
%
%\bibliographystyle{apsrev}
%\bibliography{biblio}

\end {document}